\documentclass{emulateapj}

\slugcomment{}

\shorttitle{Transition-Region Velocity Oscillations Observed by EUNIS-06}
\shortauthors{D.B. Jess et al.}

\begin{document}

\title{Transition-Region Velocity Oscillations Observed by EUNIS-06}

\author{D. B. Jess}
\affil{Astrophysics Research Centre, School of Mathematics and Physics, Queen's University, Belfast, BT7~1NN, 
Northern Ireland, U.K.}
\affil{}
\affil{NASA Goddard Space Flight Center, Solar Physics Laboratory, Code 671, Greenbelt, MD 20771, USA}
\email{djess01@qub.ac.uk}



\author{D. M. Rabin and R. J. Thomas}
\affil{NASA Goddard Space Flight Center, Solar Physics Laboratory, Code 671, Greenbelt, MD 20771, USA}

\author{J. W. Brosius}
\affil{Catholic University of America at NASA Goddard Space Flight Center, Solar Physics Laboratory, Code 671, Greenbelt, MD 20771, USA}

\and

\author{M. Mathioudakis and F. P. Keenan}
\affil{Astrophysics Research Centre, School of Mathematics and Physics, Queen's University, Belfast, BT7~1NN, 
Northern Ireland, U.K.}

\begin{abstract}

Spectroscopic measurements of AR NOAA 10871, obtained with the Extreme Ultraviolet Normal Incidence
Spectrograph (EUNIS) sounding rocket instrument on 2006 April 12, reveal velocity oscillations in the 
He~II 303.8\AA~emission line formed at $T~\approx~5~\times~10^{4}$~K. The oscillations appear to arise 
in a bright active-region loop arcade about 25$''$~wide which crosses the EUNIS slit. The period of these 
transition region oscillations is 
26~$\pm$~4~s, coupled with a velocity amplitude of $\pm~10$ km/s, detected over 4 complete cycles. Similar oscillations
are observed in lines formed at temperatures up to $T \approx 4 \times 10^5$~K, but we find no evidence for the 
coupling of these velocity oscillations with corresponding phenomena in the corona. We interpret the detected 
oscillations as originating from an almost purely adiabatic plasma, and infer that they are generated 
by the resonant transmission of MHD waves through the lower active-region atmospheres. Through use of seismological 
techniques, we establish that the observed velocity oscillations display wave properties most characteristic of 
fast-body global sausage modes.

\end{abstract}


\section{Introduction}
\label{intro}

The discovery of solar oscillations in the 1960s (Leighton~1960) and the identification of individual wave modes 
enabled accurate information on the properties of the solar interior to be established. These discoveries thereafter developed 
into the prominent research area of helioseismology (Pomerantz~et~al.~1985; Duvall~\& Harvey~1984; Scherrer~et~al.~1982). 
The discovery of oscillations in coronal structures raised the prospect of using 
these oscillations as a diagnostic tool to establish coronal loop parameters. 
The field of coronal seismology subsequently emerged (Roberts~2000; Aschwanden~et~al.~2004; Nakariakov~et~al.~2004; Ofman~2007). 
The identification of waves with wavelengths shorter than the size of typical coronal structures imposes 
great demands on current spectrocscopic and imaging instruments. The contribution of high frequency waves to the 
heating of coronal structures may be significant (Porter et al. 1994). However, current space-borne instruments lack the 
necessary spatial, spectral and temporal resolution required to probe such small-scale structures. 
Extensive analyses of photospheric oscillations has shown evidence for high-frequency, reconnection-driven acoustic wave trains 
(Jess~et~al.~2007b). High-cadence observations of the upper solar atmosphere are much more scarce, primarily due to telemetry 
restrictions imposed by space-borne experiments. 
One such study was carried out by Williams~et~al.~\cite{Wil01} who detected a 6~s intensity oscillation associated with an 
active region coronal loop. This oscillation was subsequently interpreted as a fast-mode, impulsively generated, magneto-acoustic 
wave (Williams~et~al.~2002). 

\begin{figure*}
\epsscale{0.8}
\plotone{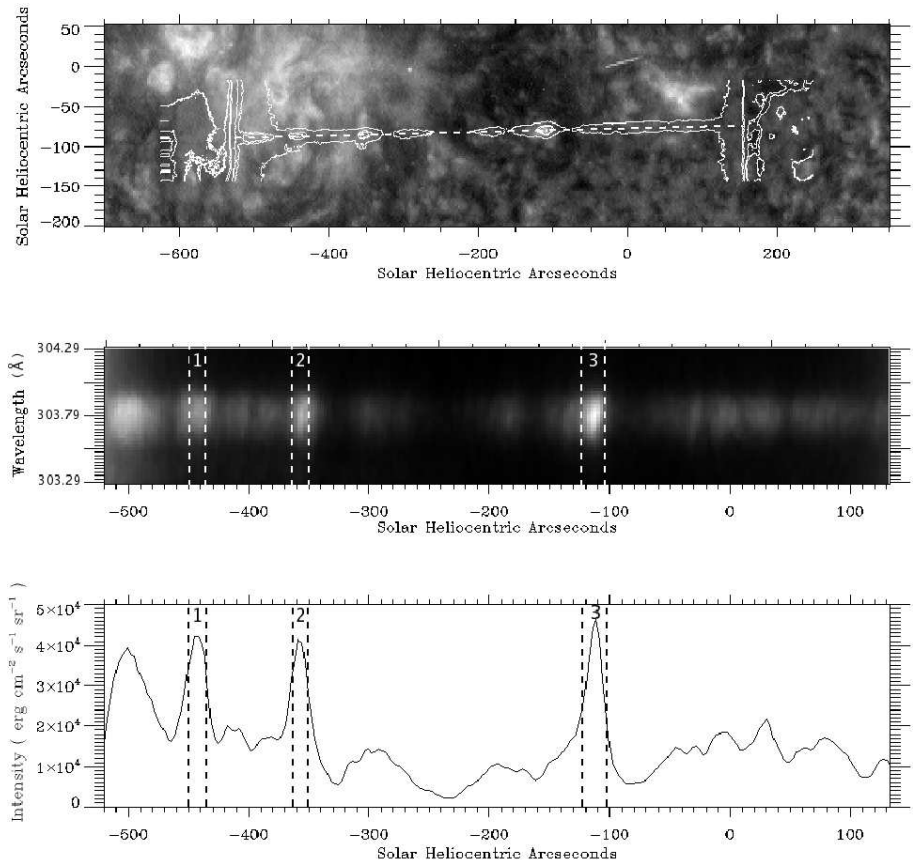}
\caption{The top panel shows the EUNIS LW~slit position, at the start of the ``stare'' sequence, after coalignment with 
{\sc{eit}}'s 304\AA~image. The middle panel shows a snapshot of the LW detector, cropped to only include contributions from the 
He~II emission line, as a function of position along the slit. The lower panel plots the integrated He~II intensity 
as a function of position along the slit. Both the middle and lower panels are overplotted with vertical 
dashed lines indicating the boundaries of three bright transition-region features, outlined in \S~\ref{analy}. Velocity 
oscillations were detected in Feature~2.
\label{fig1}}
\end{figure*}

The transition region, with a very dynamic nature over an inherently small depth, provides a regime to fully explore the 
potential coupling of underlying wave trains to those found in the corona (Doschek~2006). 
Recent work by G\"{o}m\"{o}ry~et~al.~\cite{Gom06} discusses the detection of transition region velocity oscillations with periodicities 
in the range 250--450~s and proposes this as evidence of downward propagating waves. Doyle~et~al.~\cite{Doy98} found transition region 
intensity oscillations with periods ranging from 200--500~s, while Hansteen~et~al.~\cite{Han00} report both intensity and velocity 
oscillations in the transition region with periods ranging from 100--300~s. Additionally, De~Pontieu~et~al.~\cite{DeP05} find, through 
use of numerical models, that oscillations generated in the photosphere can propagate into the corona providing they are guided 
along an inclined magnetic flux tube. Although there have been many detections of transition-region oscillations, space-borne 
instrumentation lacks the temporal resolution required to probe waves with frequencies above 33~mHz in this part of the solar 
atmosphere. The reason for this may be the lack of instruments available which are sensitive enough to obtain good count 
statistics, while maintaining short exposure times necessary for high-cadence observations. In order to probe physical processes 
occurring in the transition region on time scales less than 1~min, it is imperative to acquire data at the highest possible cadence 
using a sensitive camera system. 

Here we report velocity oscillations detected in the transition region using an Extreme Ultraviolet Normal Incidence
Spectrograph (EUNIS) sounding rocket instrument developed at NASA's Goddard Space Flight Center. In \S~2 we provide a brief outline  
of the observations, while in \S~3 we discuss the methodologies used during 
the analysis of the data and the search for reliable oscillatory signatures. A discussion of our results in the context of 
velocity oscillations is in \S~4, and finally, our concluding remarks are given in \S~5.

\section{Observations}
\label{observations}

EUNIS was launched from White Sands Missile Range, New Mexico, at 1810~UT on 2006 April~12. It achieved a maximum 
altitude of 313~km and obtained solar spectra and images between 1812 and 1818~UT. EUNIS observed NOAA AR~10871 
(S07E28) and AR~10870 (S08W02), the quiet area between them, and a southern-hemisphere coronal bright point within 
the quiet area. For the present study, we are concerned with 75 solar exposures taken during a ``stare'' (fixed pointing) 
mode near the start of the flight, with constant 1.024~s durations. Each of the data frames consist of imaged EUV spectra along a 
slit 660$''$ long, together with spectroheliograms (called lobe images) formed by 200$''$ wide areas 
beyond the two slit ends which are used for spatial registration.  For this analysis, we will focus on the long-wave [LW] channel, 
where the spectral coverage ranges from 300\AA~to 370\AA, with an associated cadence of~$\approx~2$~s. Further observations in 
scanning (spectroheliogram) mode are not discussed 
here. Additional details regarding the EUNIS instrument can be found in Brosius~et~al.~\cite{Bro07}.

\section{Data Analysis}
\label{analy}

Processing the EUNIS images was performed by subtracting an average dark frame obtained in flight, debiasing 
each detector row with a linear fit to the unilluminated ends, correcting for non-linear detector response, and applying 
a flat-field image obtained in the laboratory after the flight. Radiometric calibration, carried out post-flight at 
Rutherford Appleton Laboratory, and corrections for atmospheric extinction, are also included. Brosius~et~al.~\cite{Bro08} discuss 
validation of the EUNIS relative and absolute radiometric calibration.

\begin{figure}
\epsscale{1.0}
\plotone{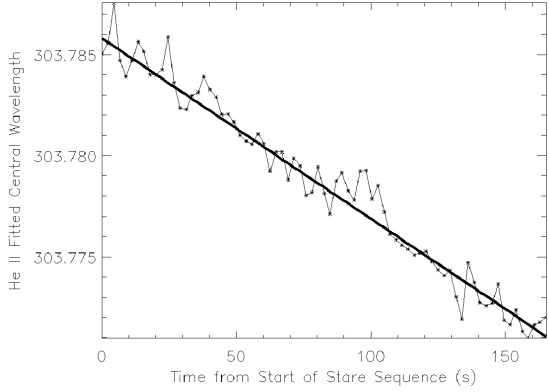}
\caption{The slit-averaged wavelength (star symbols) for each of the 75 one-second stare exposures, plotted 
as a function of time from the start of the ``stare'' sequence. For each exposure, the slit-averaged wavelength was derived 
by taking the average of 
the fitted centroid wavelength for each spatial pixel along the slit, as outlined in \S~\ref{analy}. Notice the decrease 
in wavelength, with exposure number, believed to be caused by thermal effects within the EUNIS payload. Overplotted on 
the diagram, using a solid, bold line, is a least-squares fit to the data, which is used to remove this systematic 
wavelength shift. 
\label{wave_drift}}
\end{figure}

Comparison of EUNIS He~II 303.8\AA~lobe images with a full Sun He~II image obtained at 1812 UT with {\sc{soho}}'s 
{\sc{eit}} reveals that the EUNIS~LW slit was tilted about 1.5$^{\circ}$ counterclockwise from the solar E-W axis, and that this 
angle remained 
stable in time. The EUNIS~LW field of view was initially centered at (x~=~-193$''$, y~=~-83$''$), but its pointing drifted at 
a steady rate of about 0.04$''$~s$^{-1}$ in both the x- and y-directions, which corresponds to a net drift in each dimension 
of about 6$''$ during the 167~s spent in this mode. Due to the drift in the solar x-direction being along the slit axis, we 
were able to remove this spatial shift through co-alignment techniques. Implementing Fourier cross-correlation routines as well 
as squared mean absolute deviations, we were able to provide sub-pixel co-alignment accuracy. However, it must be noted that 
sub-pixel image shifting was not implemented due to the substantial interpolation errors which may accompany this 
technique. Drift of the EUNIS~LW field of view in the solar y-direction was not compensated for since the overall drift is 
greater than the $\approx~2''$ slit-width.

After co-alignment of the EUNIS LW spectra, we averaged over three spatial pixels for a net spatial pixel size of $2.8''$.~Line fits 
were then made using Gaussian profiles on linear backgrounds, resulting in values of the centroid position, integrated intensity, 
and FWHM, together with their corresponding uncertainties, at each spatial location along the slit and for each EUNIS exposure. 
Figure 1 shows the pointing of the EUNIS LW slit near the beginning of ``stare'' sequence, along with the simultaneous He II slit 
image and corresponding line intensity as a function of position along the slit. Active Region NOAA 10871, which is the subject of 
this paper, is located towards the left-hand side of Figure 1. The EUNIS spectral resolution for this first flight was 
$\approx 200$~m\AA~FWHM in the LW channel; still, the centroids of strong lines, such as with He II reported here, can be 
determined with a statistical uncertainty of $\pm 2$~km/s or better (Brosius et al. 2007).

Since EUNIS does not provide an absolute wavelength scale, velocities reported here are relative to average values measured across 
individual solar features. For example, Figure~\ref{wave_drift} shows the slit-averaged He II central wavelength as a function of time 
from the 
start of the EUNIS stare sequence, and reveals a systematic shift toward shorter wavelength. The wavelength drift is modest, less than 
1~pixel over the whole flight, but real. Its characteristics are identical in all observed LW lines, and does not depend on the solar 
feature being observed. These facts, coupled with the shape of the drift with time, suggest it is due to thermal changes within the 
payload resulting in a tilt of the grating within its dispersion plane. To 
remove this effect, a least-squares linear fit is performed, after which the resulting trend is removed from the data leaving pure 
deviations relative to the average value measured across the solar feature. Portions of the He II slit image were analysed based on 
their relative magnitudes. Figure~1 shows the isolation of three distinct and intense features corresponding to a coronal bright 
point and two enhancements in AR~NOAA~10871.

Relative He~II velocities as a function of time are generated for each of the three features 
shown in Figure 1. Since the spatial averaging of three pixels has already been performed 
to improve the signal-to-noise ratio, no additional averaging was required. Thus, spatially resolved 
velocities are determined for each solar feature, allowing for more precise diagnostics. 
Before temporal analysis of the dataset can be implemented, timing corrections must be 
considered. During the stare mode investigated here, all exposure durations were exactly 
the same, namely 1.024~s. However, time-intervals between a small number of exposures varied somewhat. 
To compensate for this timing irregularity, all line-fit parameters were 
interpolated onto a constant-cadence time series with linear interpolation performed between data points. During this interpolation 
procedure approximately 15\% of all data points required a timing correction. 
This provides the necessary platform for temporal studies to commence.

All spatially-resolved He~II velocity measurements were passed into Fast Fourier Transform (FFT) and wavelet analysis routines. 
While a FFT searches for periodic signatures by decomposing the input signal into infinite length sinusoidal wavetrains using a 
basic exponential function, wavelet analysis utilizes a time localised oscillatory function continuous in both frequency and 
time (Bloomfield~et~al.~2004) and is therefore highly suited in the search for transient oscillations. The wavelet chosen for 
this study is known as a Morlet wavelet and is the modulation of a sinusoid by a Gaussian envelope (Torrence \&~Compo~1998). 
Strict criteria were implemented during wavelet analysis to insure that oscillatory signatures correspond to real periodicities. 
The first is a test against spurious detections of power that may be due to Poisson noise, where the input time series is 
assumed to be normally distributed (consistent with photon noise) and following a $\chi^{2}$ distribution with two degrees 
of freedom. A 99\% confidence level is calculated by multiplying the power in the background spectrum by the values of 
$\chi^{2}$ corresponding to the 99th percentile of the distribution (Torrence \&~Compo~1998, Mathioudakis~et~al.~2003).

\begin{figure*}
\epsscale{0.8}
\plotone{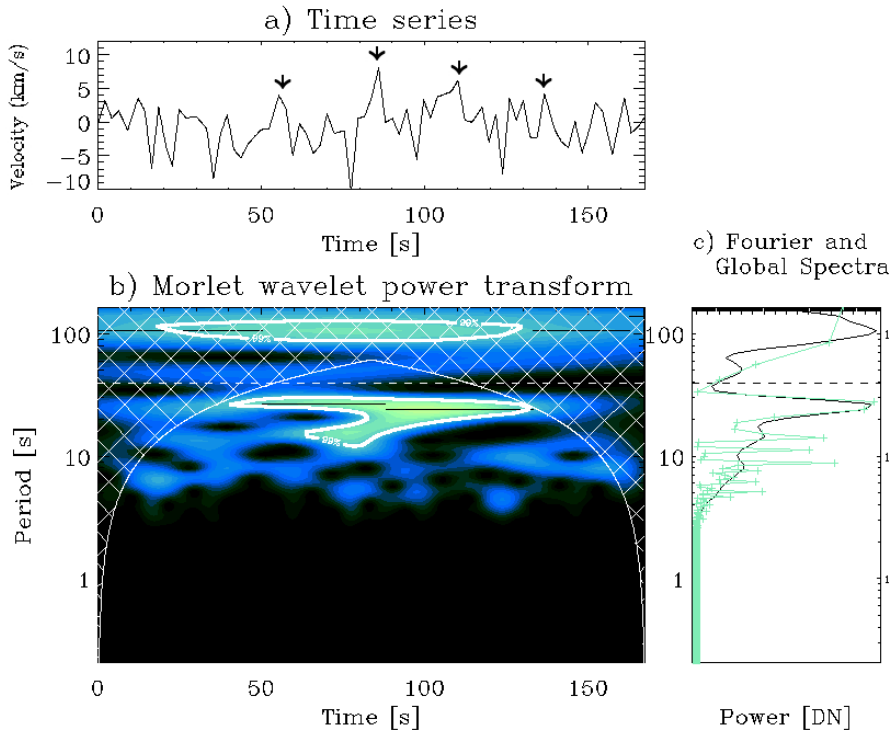}
\caption{Wavelet diagram showing the detection of a 26~$\pm$~4~s periodic velocity oscillation in He~II, over four complete cycles, 
originating from Feature~2 (see Fig.~\ref{fig1}). The original He~II velocity time series is plotted in a) with arrows indicating 
the peaks of the oscillation. The wavelet power transform along with locations where detected power is at, or 
above, the 99\% confidence level are contained within the contours in b). Plot c) shows the summation of the wavelet power 
transform over time (full line) and the Fast Fourier power spectrum (crosses) over time, plotted as a function of period. 
Both methods have detected a well pronounced 26~s oscillation. The cone of influence (COI), cross-hatched area in the plot, 
defines an area in the wavelet diagram where edge effects become important and as such any frequencies inside the COI are 
disregarded. Periods above the horizontal line (dashed) fall within the COI.
\label{wavelet1}}
\end{figure*}

The second criterion applied is a comparison of the input time series with a large number (1500) of randomized time-series 
with an identical distribution of counts. The probability, $p$, of detecting non-periodic power is calculated 
for the peak power at each timestep by comparing the value of power found in the input time series with the number of times 
that the power transform of the randomized series produces a peak of equal or greater power. A percentage confidence is attributed 
to the peak power at every time step in the wavelet transform by $(1 - p) \times 100$, such that a high value of $p$ implies 
that there is no periodic signal in the data, while a low value suggests that the detected periodicity is real 
(see Banerjee~et~al.~2001). 

Our final wavelet criterion is the exclusion of oscillations which last, in duration, less than $1.41$ cycles. This is consistent 
with the decorrelation time defined by Torrence~\& Compo~\cite{Tor98}. One can distinguish between a spike in the data and a 
harmonious periodic component at the equivalent Fourier frequency by comparing the width of a peak in the wavelet power spectrum 
with the decorrelation time. From this, the oscillation lifetime at the period of each power maximum is defined as the interval 
of time from when the power supersedes 95\% significance to when it subsequently dips 
below 95\% significance (McAteer~et~al.~2004). The lifetime was then divided by the period to give a lifetime in terms 
of complete cycles (Ireland~et~al.~1999). Any oscillations which last for less than this minimum duration were discarded as they 
may have simply been a spike in the time series.

Wavelet power, for a variety of oscillatory periods which lay above the 95\% significance level, was saved as the output of 
this analysis.

\section{Results and Discussion}
\label{results}

In Figure~\ref{wavelet1}, it is clear to see a strong 26~s oscillatory signal, coupled with an amplitude of $\pm$10~km/s, 
lasting in excess of four complete cycles. Through examination of the FWHM of corresponding Fourier and global spectra, in 
addition to the boundaries of the 99\% confidence-level contours, we can associate an error of $\pm$~4~s to the detected periodic 
signal. This oscillatory frequency is much higher than previously detected and originates from the most westerly part of the 
active region (Feature~2). Similar analysis of the intensity variation of Feature~2 reveals no intensity oscillations at this 
location. Through closer examination of the {\sc{eit}}~304\AA~image, 
the location of strong oscillatory power originates from a particularly bright part of AR~NOAA~10871. 
Due to the EUNIS instrument providing spatially-resolved spectra, pixels of the EUNIS slit corresponding to this bright 
feature (eight in total) were analysed in full. Each pixel, corresponding to a spatial scale of 2.8$''$ (as outlined in 
\S~\ref{analy}), provides high oscillatory power at a 26~s periodicity. Furthermore, no phase difference between neighboring 
pixels was found indicating that this oscillatory phenomena is of substantial physical size (in excess of 20$''$). The fact that the 
signal remains coherent during pointing drifts perpendicular to the slit-direction indicates a size of at least 6$''$ in 
that dimension as well. Additional 
transition region lines were also studied, consisting of the Mg~VI~349.2\AA~($T~\approx~4.0\times10^{5}$~K), Ne~IV~358.7\AA~
($T~\approx~1.6\times10^{5}$~K) and Ne~V~359.4\AA~($T~\approx~3.2\times10^{5}$~K) emission lines. All of these emission lines 
corroborate the detection of the 26~s velocity oscillation in Feature~2. However, due to these lines being much weaker when compared 
to He~II, the resulting reduction in signal-to-noise means that the confidence levels associated with these detections are reduced. 
Nevertheless, a strong 26~s oscillatory signal over four complete cycles, with confidence levels exceeding 85\% in each instance, 
is observed in transition region emission lines formed at temperatures up to $4 \times 10^5$~K.

\begin{figure}
\epsscale{1.1}
\plotone{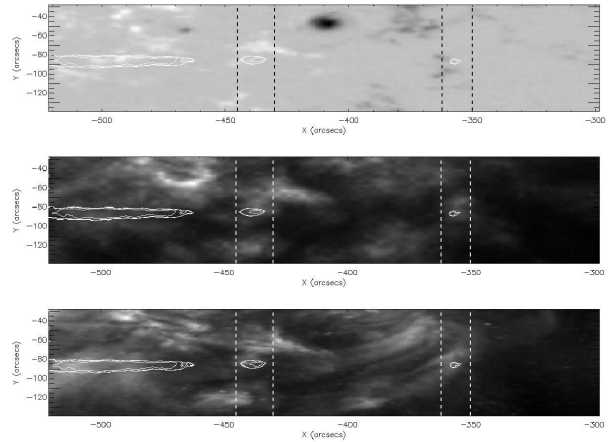}
\caption{Top panel shows an {\sc{mdi}} magnetogram, the middle panel shows an {\sc{eit}}~304\AA~transition~region image, and the 
bottom panel 
shows a {\sc{trace}}~195\AA~coronal image. Each image was taken around the time of the EUNIS launch, and therefore acts as a good 
two-dimensional reference to compare with the one-dimensional, spectroscopic results of EUNIS. Overplotted in each diagram 
are the vertical dashed lines corresponding to the boundaries of Features~1 and~2 (\S~\ref{analy} and Fig.~\ref{fig1}), as well 
as He~II intensity contours from the EUNIS~LW~slit. Notice how Features~1 and~2 are located very close to regions of opposite 
magnetic polarity.
\label{mdi_eit_trace}}
\end{figure}

Due to the full-disk nature of {\sc{eit}} and {\sc{mdi}} images, these can be accurately co-aligned by forcing the solar limbs to 
agree (L\'{o}pez Fuentes~et~al.~2006). Through multiple coalignments an uncertainty between {\sc{eit}} and {\sc{mdi}} images 
converges to approximately 0.5 of an {\sc{eit}} pixel, or 1$''$. This is an intermediary step which allows subsequent co-alignment of 
{\sc{trace}} images to {\sc{mdi}} magnetograms. Upon 
inspection of a co-aligned {\sc{trace}}~195\AA~image, it is clear that the oscillatory phenomena is generated in the immediate vicinity 
of a coronal loop structure (Fig.~\ref{mdi_eit_trace}). This, 
coupled with the strong underlying photospheric magnetic-field (top panel of Fig.~\ref{mdi_eit_trace}), 
suggest that this could be an observational signature of a propagating MHD wave. Highly magnetic structures, 
such as coronal loops, can act as superior wave guides (Edwin~\& Roberts~1983), which allow the efficient propagation of MHD waves. 
To search for upward propagating waves, we perform the same rigorous analysis on co-spatial coronal emission lines. For this, we 
choose the strong Mg~IX~368.1\AA~($T~\approx~9.5\times10^{5}$~K), Fe~XIV~334.2\AA~($T~\approx~2.0~\times~10^{6}$~K), and 
Fe~XVI~335.4/360.8\AA~($T~\approx~2.5~\times~10^{6}$~K) emission lines. Due to structures at one level in the solar atmosphere 
not always overlying the structures to which they are connected to at other levels in the atmosphere, we included additional EUNIS pixels 
during our search for coronal oscillations. Due to the EUNIS slit being one-dimensional, we are unable to investigate any oscillatory 
behavior at 
locations North and South of the EUNIS pointing. However, we analyse pixels lying to the East and West of the detected He II 
oscillations. We incorporate an extra four pixels in each direction into our investigation totalling an additional 22.4$''$. 
After searching for simultaneous oscillations in these 
lines, no velocity perturbations could be found in the corona. Additionally, no periodic signatures in coronal 
emission intensities could be found implying that our detected transition-region oscillations are not propagating out into 
the corona.

To check the validity, and reliability, of our detected He~II velocity oscillation, we examine the results of analysing two 
additional features mentioned in \S~\ref{analy}. One feature corresponds to the central portion of AR~NOAA~10871 (Feature~1), 
while the second coincides with the coronal bright point found to the west of the active region (Feature~3). Examining these 
two features in detail reveals no signatures of He~II velocity oscillations. It is interesting to note that Feature~1 
overlies a region of opposite magnetic polarity to that of Feature~2 (Fig.~\ref{mdi_eit_trace}). Additionally, 
the {\sc{mdi}} magnetogram shows that the magnetic field is stronger for Feature~2 than for 
Feature~1. In accordance with the theory of Edwin~\& Roberts~\cite{Edw83}, this may imply that the magnetic structuring around 
Feature~2 may act as a better waveguide than the magnetic field found 
surrounding Feature~1. Additionally, the velocity fluctuations of the He~II line are consistent with the relative Doppler velocities 
reported by Brosius~et~al.~\cite{Bro07}.

Such a scenario leaves open the issue of what causes these transition region velocity oscillations. De~Pontieu~et~al.~\cite{DeP03} 
state there is a general correspondence between p-modes and upper transition region oscillations, in duration, periods, and locations
of oscillatory power. Priest~\cite{Pri82} explains that the superposition of p-mode oscillations usually results in wavetrains 
containing 4 or 5 cycles, after being emitted from the photosphere. This is corroborated by De~Pontieu~et~al.~\cite{DeP03} who find 
transition region oscillations typically consisting of wavetrains 3--7 cycles long. The cycle lifetimes stated by 
De~Pontieu~et~al.~\cite{DeP03} and Priest~\cite{Pri82} match closely our findings, however, these pressure driven modes cannot 
explain the velocity oscillations presented in this work. Traditionally, transition region velocity oscillations, surrounding an 
active region, have been interpreted as the 
signature of resonant transmission of chromospheric umbral oscillations (Thomas~et~al.~1987). Previously detected transition region 
velocity oscillations have associated periods of 129--173~s (Gurman~et~al.~1982) and directly corroborate the work of 
Thomas~et~al.~\cite{Tho87}. 

Since there are no low-temperature lines in the EUV bandpasses of EUNIS, it provides no direct information about chromospheric 
plasmas. Therefore, we are unable to provide a quantitative comparison to the work of Gurman~et~al.~\cite{Gur82}, who detect transition 
region and chromospheric velocity oscillation interplay in highly magnetic structures. Nevertheless, the fact that we detect pure 
velocity oscillations, with no intensity variation, implies that we are dealing with an almost purely adiabatic plasma 
(Rendtel~et~al.~2003). Models of coupled resonators (\v{Z}ug\v{z}da~et~al.~1983, 1987; Settele~et~al.~1999), which explain the resonant 
transmission of MHD waves through active-region atmospheres, require regions of strong wave reflection, e.g. from the 
steep temperature gradient found within the transition region. Consequently, according to Rendtel~et~al.~\cite{Ren03}, 
velocity oscillations may be observed in the EUV lines of active regions as a result of resonant transmission. If 
intensity oscillations were found to accompany 
transition region velocity waves, this would hint at a non-adiabatic system consisting of isothermal waves (Staude~et~al.~1985). 
Such a distinction is important, as it indicates that the solar transition region plasma for Feature~2 is adiabatic in nature, and 
experiences strong, non-stationary behaviour over the duration of the EUNIS stare mode. Therefore, through 
comparison of our results to those of Gurman~et~al.~\cite{Gur82}, Rendtel~et~al.~\cite{Ren03} and Thomas~et~al.~\cite{Tho87}, 
we can infer that our detected transition region velocity oscillations are similar in both amplitude and location to those 
found interacting with chromospheric plasma (Brynildsen~et~al.~1999). We find the oscillations propagating in a concentrated magnetic 
field region with much higher frequencies than reported to date.

Of particular importance is the ability to place the observed He~II oscillation onto a dispersion diagram (Edwin~\&~Roberts~1983). 
From this diagram, it is possible to infer the exact mode of propagation in relation to fast- or slow-body waves. 
Several kinds of wave modes have been directly observed with the {\sc{trace}} and {\sc{soho}} 
spacecraft, as well as with ground-based telescopes (Aschwanden~et~al~1999, Wang~et~al.~2002). It is of 
particular interest that these wave modes, through use of MHD seismology techniques, 
can be deduced from simple plasma parameters such as electron density and magnetic-field strength 
(Nakariakov~et~al.~1999, Nakariakov~\& Ofman~2001). One particular mode, the sausage mode, has been typically used to interpret 
short periodicities of less than one minute. The sausage mode is based on an assumption whereby 
the period of this mode is determined by the ratio of the cross-sectional radius to the Alfv\'{e}n speed inside the 
waveguide (Meerson et al. 1978). The global sausage mode (GSM) is a fast-MHD mode which can efficiently modulate the plasma 
density and magnetic field strength. It is also one of the principal modes that can be excited in active regions providing 
the magnetic waveguide is sufficiently thick and dense. In order to assess whether the 
GSM can be supported in the transition region of AR~NOAA~10871, we require information on the plasma temperature, 
electron density and magnetic field strength.

For this assessment, we choose a plasma temperature of $T~\approx~5~\times~10^{4}$~K (the formation temperature of He~II), an 
internal-waveguide electron density of 
$n_{e}^{i}~\approx~10^{11.5}$~cm$^{-3}$, an external-waveguide electron density of $n_{e}^{e}~\approx~10^{10.7}$~cm$^{-3}$ and 
a magnetic field strength of 25~G. These values are consistent with previously measured transition region values 
(Parnell~et~al.~2002, P\'{e}rez~et~al.~1999). The GSM exists if the dimensionless wave number, $k\alpha$, is greater than 
the cut-off value, 
\begin{equation}
k_c\alpha = j_0\sqrt{\frac{(C_{S0}^2 + C_{A0}^2)(C_{Ae}^2 - C_{T0}^2)}
{(C_{Ae}^2 -C_{A0}^2)(C_{Ae}^2 - C_{S0}^2)}}
\end{equation}
where $\alpha$ is the waveguide radius and $j_0$~=~2.4 is the first zero of the Bessel function $J_{0}(x)$, as detailed in Edwin~\& 
Roberts~\cite{Edw83}. 
$C_{A0}$ and $C_{Ae}$ are the 
internal and external Alfv\'{e}n speeds, while $C_{S0}$ and $C_{T0}$~=~$C_{S0}C_{A0}$/$(C_{S0}^{2}+C_{A0}^{2})^{1/2}$ are the 
internal sound and tube speeds, respectively. These values can be derived directly 
from our input parameters described above, and following the criteria 
outlined in Nakariakov~et~al.~\cite{Nak03}, we can determine if this mode can explain the 
oscillations presented here. We derive the associated cut-off value, $k_{c}\alpha$, to be 1.12. In order to 
see whether the observed wave train is a fast- or slow-body wave, we relate the detected periodicity, $P$, to 
the internal Alfv\'{e}n speed, whereby a fast-body wave will be present if:
\begin{equation}
\alpha > \frac{j_{0} C_{A0} P}{2 \pi} .
\end{equation}

\begin{figure}
\epsscale{1.0}
\plotone{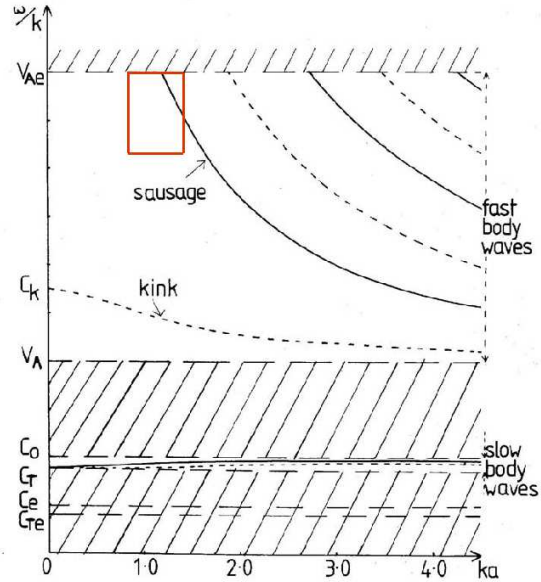}
\caption{Phase speed $\omega/k$ is shown for magneto-acoustic waves in a cylindrical 
waveguide (with radius $\alpha$) as a function of the dimensionless longitudinal wavenumber $k\alpha$ under 
the transition-region/coronal conditions $C_{Ae} > C_{A0} > C_{T0} > C_{S0}$. Solid lines show the sausage modes, while 
the dashed lines indicate kink modes (Edwin~\& Roberts~1983). The solid red box indicates the region 
on the dispersion curve in which the observed He~II velocity oscillations lie.
\label{omegak}}
\end{figure}

Substituting known values into Equation~2, we find that the minimum diameter of the waveguide 
required to channel fast-body waves is $\approx$~1800~km. We can use co-spatial coronal observations 
to provide an estimate of the diameter of the transition-region waveguide. 
By measuring the co-spatial width of coronal loops seen in the {\sc{trace}}~195\AA~images, 
we establish an average loop width of 7~{\sc{trace}}~pixels, corresponding to a diameter of $\approx$~2500~km. This measured 
width is consistent with previous {\sc{trace}} statistical loop measurements (Aschwanden~et~al.~2000, Aschwanden~\&~Nightingale~2005) 
and is greater than the minimum diameter required to channel fast-body waves. Accompanying our derived 
$k_{c}\alpha$ value, we deem that a fast-MHD wave ({\emph{i.e.}}, the global sausage mode) could explain the 
origins of the detected He~II velocity oscillation (Fig~\ref{omegak}). In addition to He~II, the warmer 
transition region lines for which periodic velocity oscillations were detected are also consistent with the global sausage mode. 
Increasing the line-formation temperature from $5~\times~10^{4}$ to $4.0\times10^{5}$~K leads to an increase in the sound and tube 
speeds, whereas the internal and external Alfv\'{e}n speeds remain unchanged. As a result, the cut-off value is increased, thus 
maintaining the validity of the GSM as the most favorable mode of propagation.

However, it must be noted that the cutoff value, $k_{c}\alpha$, 
is extremely sensitive to the assumed values of magnetic field strength and electron density, through their relationship with the 
Alfv\'{e}n speed, $C_{A}$. Furthermore, the magnetic field strength is almost always an inferred measure (Parnell~et~al.~2002) 
and exerts a stronger variation on the Alfv\'{e}n speed due to a linear dependence between them. Therefore, caution must be 
exercised when evaluating the mode of propagation. In this instance, the GSM appears to be the most favorable mode of propagation, 
although for a definitive answer, values of magnetic field strength and electron density must first be determined accurately.

\section{Concluding Remarks}
\label{conc}

We have presented direct evidence of high-frequency waves propagating in an active region, at transition region heights. We have 
shown the location of 26~s oscillatory signatures, with significance levels greater than 95\% due to the rigorous wavelet criteria 
enforced in \S~\ref{analy}. The velocity oscillations detected here originate from a large ($>$ 20$''$), bright He~II structure 
crossing the EUNIS slit on 
the westerly edge of AR~NOAA~10871. The highest temperature at which we observe this oscillation is $4 \times 10^5$~K, but we find no 
evidence for upward propagation of this wave train into the corona, and infer, 
qualitatively, that the detected oscillations could be caused by the resonant transmission interplay between the 
chromosphere and transition region. We deduce that the observed velocity oscillations exhibit signatures of a fast-MHD wave - in 
particular, that of the fast-body global sausage wave mode. 
The work of De Pontieu~et~al.~\cite{DeP03} emphasizes the coupling of transition region oscillations to those in the underlying 
photosphere. A natural extension of this work is to investigate if such a correspondence also exists in the high frequency 
domain (see Jess~et~al.~2007a for high-frequency photospheric analysis). 


\acknowledgments

DBJ is supported by a Northern Ireland Department for Employment and Learning studentship. DBJ additionally thanks NASA 
Goddard Space Flight Center for a CAST studentship -- in particular Doug Rabin, Roger Thomas and Jeff Brosius deserve special 
thanks for their endless help, support and scientific input. FPK is grateful to AWE Aldermaston for the award of a William Penney 
Fellowship. The EUNIS program is supported by the NASA Heliophysics 
Division through its Low Cost Access to Space Program in Solar and Heliospheric Physics. We thank the entire EUNIS team for 
the concerted effort that led to a successful flight. Wavelet software was provided by C. Torrence and G.P. Compo.
\footnote{Wavelet software is available at http://paos.colorado.edu/research/wavelets/.}

\clearpage

\clearpage

\clearpage

\clearpage

\clearpage

\end{document}